\documentclass[12pt,aps,prd,preprint,tightenlines,superscriptaddress,
   nofootinbib]{revtex4}
\usepackage{graphicx}
\usepackage{graphicx}

\newcommand{\PRE}[1]{{#1}} 

\newcommand{\roughly}[1]{\mathrel{\raise.3ex\hbox{$#1$\kern-0.85em
\lower1ex\hbox{$\sim$}}}}

\def\be{\begin{equation}}
\def\beq\begin{equation}
\def\ee{\end{equation}}
\def\bea{\begin{eqnarray}}
\def\eea{\end{eqnarray}}

\def\eqref#1{(\ref{#1})}

\def\UV{{\scriptscriptstyle U \kern-.1emV}}
\def\IR{{\scriptscriptstyle I\kern-.18em R}}

\newcommand\del{\partial}

\begin{document}

\preprint{UCI-TR-2016-13}

\title{\PRE{\vspace*{1.3in}}
 de Sitter Space is Unstable in Quantum Gravity
\PRE{\vspace*{0.3in}}
}

\author{Arvind Rajaraman%
\PRE{\vspace*{.4in}}
}
\affiliation{Department of Physics and Astronomy, University of California, Irvine, CA  92697 USA
\PRE{\vspace*{.5in}}
}


\begin{abstract}
  \PRE{\vspace*{.3in}}
 Graviton loop corrections to observables in de Sitter space often lead to infrared divergences.
We show that these infrared divergences  are resolved by
the spontaneous breaking of de Sitter invariance. 
\end{abstract}

\maketitle

\section{Introduction}

Quantum effects in de Sitter (dS) space have become central to particle physics and cosmology 
since the measurement of the cosmic microwave background~\cite{Bennett:2012fp} and its
interpretation as being produced from zero point fluctuations around the
vacuum~\cite{Linde:2005ht}. Much effort has been expended in trying to extend this to higher precision by including  
loop corrections to this process. 
In particular,  loop corrections coming from gravitons
 have been considered~\cite{Seery:2007wf,Dimastrogiovanni:2008af,
Tsamis:1993ub,Rajaraman:2010zx,Tsamis:1996qq,Tsamis:1996qm}, with mixed success. The major impediment is 
the presence of infrared (IR) divergences in the loop calculation.
It turns out that the graviton
propagator in de Sitter is more singular at low momenta than the
corresponding propagator in flat space; loop corrections with an internal graviton propagator then
 become divergent at low graviton momenta~\cite{Giddings:2010nc,Giddings:2010ui,Giddings:2011zd}.
In  particular, graviton corrections to the CMB are divergent, naively invalidating the successful tree level calculation.

Many approaches have been taken to  resolve this issue. We review several of
these below, before moving to our own attempt at a resolution.

(1) One general approach has been to consider loop corrections involving massless scalars. It happens to be the case 
that massless scalars in de Sitter space display the same type of infrared divergence as gravitons; indeed, the 
graviton propagator
in a suitable gauge is identical to the propagator of two massless scalars~\cite{Ford:1977in,Kirsten:1993ug}. An 
understanding of the IR divergences for a massless scalar might then be hoped to shed
light on the divergences for a graviton.

Unfortunately this turns out not to be the case. Multiple analyses using multiple approaches 
e.g. Euclidean continuation~\cite{{Rajaraman:2010xd}, Beneke:2012kn,Marolf:2010zp,Marolf:2010nz,Higuchi:2010xt}, the stochastic 
formalism~\cite{Starobinsky:1994bd,Garbrecht:2013coa,Boyanovsky:2011xn}, the dynamical RG~\cite{Burgess:2010dd,Burgess:2009bs},  
truncated Schwinger-Dyson equations~\cite{Serreau:2013koa,Serreau:2013eoa,Gautier:2013aoa,Serreau:2013psa,Garbrecht:2011gu,Rajaraman:2015dta}
and others~\cite{Boyanovsky:2005sh,Riotto:2008mv,Herranen:2013raa,Jatkar:2011ju,Seery:2007we,Youssef:2013by}
have conclusively 
resolved the physics of a massless scalar in de Sitter space. The scalar develops
a dynamical mass; for example, an apparently massless scalar with a $\lambda\phi^4$
interaction develops a dynamical mass proportional
to $\sqrt{\lambda}$.
The IR divergence is then absent, but the perturbation expansion becomes
an expansion in $\sqrt{\lambda}$ rather than $\lambda$.

Now this resolution, which is perfectly satisfactory for scalars, is no
help at all for the graviton case. The gauge invariance of gravitational perturbations around de Sitter space  
precludes the development
of a mass, dynamical or otherwise, for the graviton. Any corrections to the graviton
propagator which preserve de Sitter invariance are necessarily suppressed by powers of the invariant de Sitter momentum,
and are therefore irrelevant for solving the problem of the infrared divergences. 
We must therefore seek the resolution elsewhere.

(2) Since the primary difference between the scalar and the graviton is the existence of
a gauge symmetry, one may wonder if the gauge symmetry itself removes the
divergence. This possibility has been developed by \cite{Higuchi:2011vw}, who argue that the IR
divergence is in fact a gauge artifact (see also \cite{Faizal:2016fri}).  The evidence for this comes from an analysis of
the graviton propagator. The IR divergence translates into a growth of the propagator at
large separations; the authors show that this growth can be canceled by a gauge transformation, albeit one that 
also grows at large distances. This suggests that a suitable
limiting procedure can remove the infrared divergences. 

The problems with this approach have been elaborated in 
detail in \cite{Miao:2011ng}. In addition to the points
raised there, we note that if the IR divergence was indeed a gauge artifact,
then loop corrections to gauge invariant quantities would not have this divergence. This is explicitly contradicted by 
calculations, and hence it does not
seem possible to gauge away the divergence.

(3) Woodard and collaborators have suggested that the correct approach is to modify the 
graviton propagator ab initio (a few of these papers are \cite{Tsamis:1996qk,Miao:2011fc,Glavan:2015ura}). Since the 
de Sitter invariant propagator
 has the issues described above, they use a non invariant propagator (which satisfies the 
same differential equation as the de Sitter propagator, but does not have the full de Sitter invariance).  
  This is clearly an explicit breaking of the symmetry; for example, the analysis \cite{Glavan:2015ura} shows that 
an explicit non-invariant counterterm is required to cancel a loop divergence. 
Whether such a term is allowed in a consistent theory is unclear--- explicit symmetry breaking
in a gauge theory typically leads to violations of renormalizability and unitarity. It is also
hard to see how the standard sum over metrics in the path integral would lead to such a 
propagator. Finally, at least in string theory,
calculations of the  effective action show no sign of counterterms breaking gauge
invariance (see e.g. \cite{Rajaraman:2005up}).
Lacking a quantum theory of gravity in de Sitter space, we cannot rule out the possibility that
this is a consistent approach, but clearly other approaches should be considered.

(4) If it is not possible to have a propagator which is de Sitter invariant, and 
if we do not wish to explicitly break the symmetry, then that leaves only one option
--- de Sitter invariance must be {\it spontaneously} broken. We will explore this possibility
 in this paper, and argue that this is indeed the appropriate resolution.

The idea that de Sitter space is unstable has been forcefully advocated by
Polyakov ~\cite{Polyakov:2007mm,Polyakov:2009nq,Polyakov:2012uc,Krotov:2010ma} (for related ideas, 
see \cite{Bander:2010zb,Anderson:2013zia,Anderson:2013ila,Bros:2008sq,Ho:2015bua,Akhmedov:2013xka,Antoniadis:1995fc,Antoniadis2}). In these papers, 
Polyakov has argued
that a loop calculation in scalar field theory in de Sitter space already shows
IR divergences which can be interpreted as 
catastrophic particle production, leading to a decay of de Sitter. 
However, other calculations show no signs of an instability 
e.g \cite{Jatkar:2011ju}. 
Furthermore, it has 
been argued \cite{Higuchi:2010xt} that the scalar field theory
can be formulated in an Euclideanized version of the theory,
in which the finiteness properties can be proven \cite{Marolf:2010nz}.

The difference between these results appears to be due
to the choice of formalism.
While Polyakov has argued that one should use  the in-out formalism, which calculates the transition 
amplitudes from the earliest times to the latest times,
most calculations use 
the in-in (or Schwinger-Keldysh or CTP)
formalism  \cite{Weinberg:2005vy,Weinberg:2006ac,vanderMeulen:2007ah,Sloth:2006az,Sloth:2006nu,Senatore:2009cf,Kumar:2009ge}
which calculates transitions effectively from the earliest times to a finite time in 
de Sitter in a particular choice of time slicing (we shall elaborate further on this below). 
This may suggest that the instability found by Polyakov in scalar field theory
is an artifact of
the choice of the in-out formalism.
This is worrying because perturbation theory in the in-out formalism is known to fail
for other theories in de Sitter space. For example,  the exact solution for
the massive scalar propagator is known, but cannot be reproduced in the in-out formalism using the 
mass term as a perturbation~\cite{Higuchi:2009ew}.
This failure can be attributed to the out-vacuum not being close to the in-vacuum due to 
particle production in de Sitter 
space~\cite{Higuchi:2008tn}. The stability of scalar field theory in de Sitter is
therefore an open issue.

In this paper, we shall apply the in-in formalism, not to scalar field theory, but to gravitational 
perturbations around de Sitter 
space. As we have already argued, the infrared divergences point to a spontaneous breakdown of the 
de Sitter symmetry.
We show that indeed loop corrections lead to an instability of the de Sitter metric and a deformation 
of the metric. We therefore 
conclude that the instability argued for by Polyakov definitely does occur when gravitational 
perturbations are considered.

We note that  results similar to ours have also been arrived at in the papers
\cite{Giddings:2010nc,Giddings:2010ui,Giddings:2011zd}. In these papers the fluctuations
of geodesics are considered. The authors find that (to quote from the
abstract of \cite{Giddings:2011zd})
 ''metric perturbations produce significant and growing corrections to the lengths of such 
geodesics..
These become large, signaling breakdown of a perturbative description of the geometry via such observables,
and consistent with perturbative instability of de Sitter space''. This is qualitatively
similar to our result, but appears to disagree quantitatively, because we find a perturbatively
calculable corrections to the metric, while the authors of \cite{Giddings:2011zd} appear to
find a nonperturbative effect. It would
be very interesting to see if these results are consistent with each other. 
 
In the following section, we set out some well-known properties of de Sitter space and
gravitational perturbations around it (this section also serves to establish our notation
and to review the in-in formalism).
We show that indeed the gravitational perturbations  satisfy the same equations
as a massless scalar, and that this leads to divergences in loop calculations.

We then proceed, in the subsequent section, to regulate this divergence. We do this by
deforming the metric slightly away from de Sitter (we choose a deformation which is
spatially homogeneous and rotationally invariant). As one might expect, this modifies the equations
for the gravitational perturbations slightly, and they no longer satisfy the
same equation as a massless scalar. In such a background, the loop
calculations are well defined, and the effective action can be computed in the usual way.
Crucially, if the deformation is small, traces of the IR divergence are still visible
in the enhancement of certain correlation functions. We explicitly show that
if the deformation is parametrized by a small parameter $\epsilon$, then
the propagator of the gravitational perturbations is enhanced by a factor ${1\over \epsilon}$.
The effective action and the effective equations of motion then have terms
which are enhanced by this factor.

We then look for solutions to the effective equations of motion. Classically, the only
solution to the equations of motion is de Sitter space. This is seen by the fact that we find a tadpole
for perturbations around the deformed space. These tadpoles vanish classically 
only when $\epsilon=0$; that is, for de Sitter space. However, we find that the quantum effective equations of motion
has new tadpoles, some of which are even enhanced by a factor ${1\over \epsilon}$. The tadpole cancellation
now occurs when $\epsilon$ is nonzero. That is, de Sitter space, corresponding to $\epsilon=0$, is {\it not} a
 solution to the quantum corrected equations of motion. This explicitly shows that the de Sitter symmetry is spontaneously broken.

As part of this calculation, we are also able to estimate the value of $\epsilon$ at which the geometry is 
stabilized in the quantum theory. We find that $\epsilon$ scales 
as $\sqrt{\kappa}$, where $\kappa=8\pi G$ is the coupling constant in gravity. The propagators are then 
enhanced by a factor ${1\over \epsilon}$ leading to a perturbation expansion in powers of $\sqrt{\kappa}$ rather
than the expected $\kappa$. This behavior has similarities to the massless scalar. 

We finally close with a discussion of our results.

\section{Gravitational perturbations around de Sitter space }
\subsection{Notation and Overview}
We shall work in the mostly minus signature. Indices $\mu,\nu$ will run over 0,1,2,3 while
indices $i,j$ run over 1,2,3. The time coordinate $x^0\equiv \tau$. We follow the conventions
\bea
R^\rho_{~\sigma\mu\nu}=-\del_\mu \Gamma^\rho_{\nu\sigma}+\del_\nu \Gamma^\rho_{\mu\sigma}
-\Gamma^\rho_{\mu\lambda}\Gamma^\lambda_{\nu\sigma}+\Gamma^\rho_{\nu\lambda}\Gamma^\lambda_{\mu\sigma} 
\qquad 
R_{\mu\nu}=R^\rho_{~\sigma\mu\rho}
\eea

The Einstein action 
coupled to a cosmological constant is
\bea
{\cal L}={1\over 2\kappa}\sqrt{g}(R+2\Lambda)
\eea
where $\kappa=8\pi G$.
The equation of motion  
admits the de Sitter solution, written in Poincare patch coordinates as
\bea
ds^2={1\over H^2\tau^2}(d\tau^2-dx_idx^i)
\eea
where 
\bea
H^2={\Lambda\over 3}
\eea
Equivalently, the de Sitter metric is
\bea
\bar{g}_{00}={1\over H^2\tau^2}
\qquad
\bar{g}_{ij}=-{1\over H^2\tau^2}\delta_{ij}
\eea
Here the spatial coordinates $x^i$ run from $-\infty$ to $\infty$. The coordinate $\tau$ runs from $-\infty$
corresponding to early times, till $\tau=0$ which corresponds to late times. 

\subsection{Scalar perturbations around de Sitter and the in-in formalism}

As we have said above,  calculations in de Sitter space
need to be done using the in-in formalism.
We shall here summarize a few important details of this formalism which
are necessary for us; 
more details can be found in~\cite{Weinberg:2005vy,vanderMeulen:2007ah}. We shall use the scalar field 
theory as an example.

The in-in formalism requires us to double the number of
fields. For a scalar field, we go from $\phi $ to $\phi ^+, \phi ^-$.
 The two fields 
are constrained to be equal at a time $\tau_0$; i.e.
$\phi^+(\tau_0)=\phi^-(\tau_0)$. $t_0$ is the intermediate time at
which correlations are calculated.

The Lagrangian ${\cal L}(\phi )$ is replaced by 
${\cal L}^{in-in}={\cal L}(\phi ^+)-{\cal L}(\phi ^-)$. 
This formalism therefore has two independent 
propagators; in addition to the standard propagator, one
must introduce the Schwinger-Keldysh propagator:
\bea
F(x,y)={1\over 2}(\langle \phi(x)\phi(y)\rangle +\langle \phi(y)\phi(x)\rangle)
\eea

We can find this propagator explicitly for a free scalar. 
Such a scalar satisfies 
\bea
\del_\mu(\sqrt{g}g^{\mu\nu}\del_\nu)\phi_{m^2}+{m^2\over H^4\tau^4}\phi_{m^2}=0
\eea
We perform a Fourier transform in the spatial directions; we then find~\cite{vanderMeulen:2007ah}
\bea
\phi_{m^2}(t,\vec{k})=-{H\tau\sqrt{-\pi\tau}\over 2} H_{\nu}(-k\tau)
\eea
where $H_\nu(-k\tau)$ is a Hankel function, and 
 $\nu^2 ={9\over 4}-{m^2\over H^2}$. The Schwinger-Keldysh propagator is
then
\bea
F(k,\tau_1,\tau_2)={\pi H^2\tau_1^{3/2}\tau_2^{3/2}\over 4}Re(H_\nu(-k\tau_1)H^*_\nu(-k\tau_2))
\eea
An important special case is the massless scalar. This  sets $m^2=0, \nu={3\over 2}$. We have then
\bea
\phi_{m^2=0}(t,\vec{k})=i{H\over \sqrt{2k^3}}(1+ik\tau) e^{-ik\tau}
\eea
and
\bea
F(k,\tau_1,\tau_2)={H^2\over {2k^3}}[(1+k^2\tau_1\tau_2)\cos(k(\tau_1-\tau_2))
+k(\tau_1-\tau_2)\sin(k(\tau_1-\tau_2))]
\eea

The final important point is that for very small $k$, the massless propagator limits to 
\bea
F(k,\tau_1,\tau_2)\to {H^2\over {2k^3}}
\eea
while the massive propagator limits to
\bea
F(k,\tau_1,\tau_2)\to {H^2\over 2k^{3}}(k^2\tau_1\tau_2)^{m^2\over 3H^2}
\eea

 Note that $F(x,y)$ is not well defined for a massless scalar; the inverse Fourier transform does
not exist. This is a manifestation of the well known fact
that a massless minimally coupled scalar in de Sitter space
does not have a de Sitter invariant propagator~\cite{Ford:1977in,Kirsten:1993ug}.

\subsection{Linearized gravitational perturbations}
Gravitational fluctuations around the background metric are parametrized as
\bea
g_{\mu\nu}=\bar{g}_{\mu\nu}+h_{\mu\nu}
\eea

Up to second order, the Lagrangian for these perturbations is found to be \cite{Goroff1} 
\bea
{1\over 2\kappa H^4\tau^4}\left({1\over 8}(\bar{D}_\mu h)^2
-{1\over 4}(\bar{D}_\nu h_{AB})^2
+ {1\over 2} (\bar{D}_B h_{A}^{~B}-{1\over 2}\bar{D}_A h)^2
- {1\over 2} h^{A\nu}\bar{R}_{AF\nu B}h^{FB}+{1\over 4}\Lambda h^2\right)
\eea
where $h=\bar{g}^{\mu\nu}h_{\mu\nu}$, barred covariant derivatives are taken with respect to the background
de Sitter metric, and indices are raised and lowered with the background metric.

We can solve the corresponding equations of motion in a gauge $h_{00}=h_{0i}=0$. The 
equations with indices $0i,00$ set $\del_ih_{ij}=h=0$.
A perturbation is then characterized by the transverse traceless
polarizations. These satisfy~\cite{Woodard:2015kqa}
\bea
\del_\mu(\sqrt{\bar{g}}\bar{g}^{\mu\nu}\del_\nu(\tau^2h_{ij}(\tau,\vec{x})))
=0
\eea
It is convenient
to define
\bea \gamma_{ij}=H^2\tau^2h_{ij}
\eea
Then $\gamma_{ij}$ satisfies the same equation as for a massless scalar field.

The Schwinger-Keldysh propagator can now be determined. We will
particularly be interested in the coincident limit, where we find
\cite{Giddings:2010nc}
\bea
\langle \gamma_{ij}(x)\gamma_{kl}(x)\rangle=\int {d^3q\over (2\pi)^3}{H^2\over q^3}(1+q^2\tau^2) P_{ijkl}
\eea
where the last factor is a projection
operator
\bea
P_{ijkl}=\delta_{ik}\delta_{jl}+\delta_{il}\delta_{jk}-\delta_{ij}\delta_{kl}
+\delta_{ij}\hat{q}_k\hat{q}_l
+\delta_{kl}\hat{q}_i\hat{q}_j
-\delta_{ik}\hat{q}_j\hat{q}_l
-\delta_{il}\hat{q}_j\hat{q}_k
-\delta_{jk}\hat{q}_i\hat{q}_l
-\delta_{jl}\hat{q}_i\hat{q}_k\nonumber
\\
+\hat{q}_i\hat{q}_j\hat{q}_k\hat{q}_l
\eea
where $\hat{q}$ is the unit vector in the direction of $q$.

For a graviton (or a massless scalar) this propagator
goes as ${1\over q^3}$ at small $q$. Any loop
diagram with an internal $F$ propagator then diverges (as $\int {d^3q \over q^3}
$) when the momentum flowing through this propagator goes to zero. (Examples of 
such calculations can be found in e.g. \cite{Giddings:2010nc}.)
This makes essentially all loop calculations ill-defined, and in particular any computation
of the gravitational effective action around de Sitter is impossible.
Just as for a massless scalar, $F(x,y)$ is not
well defined.

We note that for a massive scalar, the divergent integral is
regulated by the mass and becomes of the form  $\int {d^3q \over q^{3-2\epsilon}}
$ which is finite. For the graviton, gauge invariance prevents such a mass
from being generated.

\section{Deforming Away from de Sitter}
\subsection{Gravitational Perturbations}

Since our calculation of the quantum effects around de Sitter have been 
derailed by the infrared divergences, we move slightly away from de Sitter
and attempt to calculate the effective action around a slightly deformed
metric. While we can in principle consider any deformation, we will restrict
ourselves to a deformation that preserves the spatial rotation and spatial translational
symmetry. We therefore consider a metric
of the form
\bea
g^{(ddS)}
_{00}={1\over H^2\tau^2}
\qquad
g^{(ddS)}_{ij}=-{1\over H^2\tau^2}f(\tau)\delta_{ij}\label{ddsmetric}
\eea
Here the superscript ddS stands for deformed de Sitter. 
The function $f(\tau)$
parametrizes the deformation. 
We will assume that $f(\tau)$ is close to 1.

We can now consider perturbations around this metric. This is hard to do
for a general background metric, but is facilitated here by the fact that this metric
is close to the de Sitter solution.
Accordingly, the equations of motion for the
perturbations will also be close to the de Sitter equations of
the previous section. We  continue to use the gauge $h_{00}=h_{0i}=0$.

We found in the previous section that  the transverse traceless perturbations around de Sitter
satisfy the same equation as a massless scalar. This led to the solution being singular at small
momenta, and so we are interested in whether the deformation can 
modify the equation at low momenta.
One possibility is that the deformation causes
the transverse traceless perturbations to mix with the other perturbations. This however does
not happen; the rotational symmetry requires any such mixing to come
with derivatives, and this mixing is then suppressed at low momenta.

Direct evaluation of the equations (using Mathematica~\cite{Mathematica}) shows that the transverse traceless
perturbations now have the action (this is for a perturbation with momentum
along the z-direction)
\bea
{\cal L}={H^2\over 8\kappa f^{3/2}}\left(h_{ij}^2(-2f-3\tau f'+\tau^2f'')+\tau^2(\del_zh_{ij})^2
-f\tau^2(\del_\tau h_{ij})^2\right)
\eea
Here primes indicate derivatives with respect to $\tau$.

To bring this to the form of a  scalar action, we define
\bea
\gamma_{ij}={H^2\over f^{1/4}}\tau^2h_{ij}
\eea
This field satisfies the equation of motion
\bea
{\del\over \del\tau}\left({1\over \tau^2}{\del\over \del\tau}\gamma_{ij}\right)
+{1\over f\tau^2}k^2\gamma_{ij}+{5\over 4\tau^3f}\gamma_{ij}(-2f'+tf'')=0
\eea

\subsection{A Special Case}

Let us focus on a particular case. We take $f=\tau^{\epsilon}$ where $\epsilon$ is small. 
The equation then becomes (to order $\epsilon$)
\bea
{\del\over \del\tau}\left({1\over \tau^2}{\del\over \del\tau}\gamma_{ij}\right)
+{\tau^{-\epsilon}\over \tau^2}k^2\gamma_{ij}-{15\over 4\tau^4}\epsilon\gamma_{ij}=0
\eea

The perturbations no longer satisfy the equation for a massless de Sitter scalar. In particular, a term similar to the 
scalar mass term has appeared.
We can solve this equation as we did
for a massive scalar
\bea
\gamma_{ij}(k,\tau)=-{H\tau\sqrt{-\pi\tau}\over 2} H_{\nu}(-k\tau)
\eea
with $\nu^2={9\over 4}-{15\epsilon\over 4}$. The coincident limit of the propagator
is then modified to 
\bea
\langle \gamma_{ij}(x)\gamma_{kl}(x)\rangle=\int {d^3q\over (2\pi)^3}{H^2\over q^3}(1+q^2\tau^2)
(q^2\tau^2)^{5\epsilon\over 4} P_{ijkl}\sim 
{ {H^2}\over 5\epsilon \pi^2}P_{ijkl}\label{coincident}
\eea
The coincident limit of the propagator, and in fact the propagator in general, is now finite,
in contrast to the de Sitter case. The IR divergence is regulated, and loop calculations are now well defined.
Note that the propagator is enhanced by a factor ${1\over \epsilon}$, which is
the vestige of the IR divergence.

(We note parenthetically that this is not exactly correct
because the equation is not quite that of a massive scalar ---
  there is a time dependence in the momentum dependent term. However, this can be treated as  a 
slow variation of
$k$ with time. The solution to this mode equation is then approximately the solution above
 with $k$ replaced by $k\tau^{-\epsilon/2}$. This does not
affect the low momentum behavior of the solution.)

\section{Tadpole cancellation}

We must however address another issue; the metric we are expanding around is not a 
solution to the classical equations of motion. Accordingly,  if
we were to consider the trace perturbations around the metric (\ref{ddsmetric}), 
we would find a tadpole.
Specifically consider a fluctuation around the metric (\ref{ddsmetric}) of the form 
\bea
g_{00}=g^{(ddS)}_{00}
\qquad
g_{ij}=g^{(ddS)}_{ij}+h\delta_{ij}
\eea
Classically, the equation of motion for the perturbation $h$ does not allow the solution $h=0$.
Indeed, to leading order in $(f-1)$, the perturbation $h$ satisfies the equation
(after a suitable normalization of the kinetic term)
\bea
8h-8\tau h'-4\tau^2h''={(8f'-4tf'')\over H^2\tau \sqrt{\kappa}}
\eea

This tadpole vanishes when $f$ is a constant, which  is the statement that 
classically, the only solution to Einstein's equations of the form (\ref{ddsmetric}) is the de Sitter metric.

However, this does not have to be the case at the quantum level. There is a new tadpole generated at 
one loop that contributes
to the equations. This can modify the equation for $h$. 
More precisely, there can
be terms in the action roughly of the form $hh_{ij}^2$. At one loop, we find a tadpole for $h$ when $h_{ij}^2$ is
replaced by its propagator. 

These trilinear couplings can be very complicated~\cite{Goroff1}, and it is
useful to look for simplifications. In this case, we will use the fact that we are
expanding around a metric which is close to de Sitter. As we have seen above, the propagator
for the transverse traceless polarizations then receives a enhancement, and will dominate the tadpole.
For this enhancement to occur,  no derivative can act on the propagator. This means that we
are only interested in a trilinear coupling $hh_{ij}^2$ where no
derivative acts 
on the $h_{ij}$.
It is straightforward to expand the action to find this term.

We find that this new coupling modifies the equation of motion to 
\bea
8h-8\tau h'-4\tau^2h''={(8f'-4\tau f'')\over H^2\tau \sqrt{\kappa}}+2{\sqrt{\kappa} H^2} h_{ij}^2\tau^2\label{tadpole}
\eea

At one loop, 
we should replace the term $h_{ij}^2$ by the
coincident  limit of the Schwinger-Keldysh propagator evaluated
in the metric (\ref{ddsmetric}). While we cannot solve this in general, we
can find the solution for the case
$f=\tau^\epsilon$. Here we have from eqn (\ref{coincident})
\bea
\langle h_{ij}(x)h_{ij}(x)\rangle=
{ 4\over 5H^2\tau^4\epsilon \pi^2}
\eea

The tadpole in (\ref{tadpole}) now cancels if
\bea
{\epsilon^2 }+{ 2\kappa H^2\over 15 \pi^2}=0
\eea
indicating that the quantum equations of motion are indeed solved by a metric of 
the form (\ref{ddsmetric}) with $f=\tau^\epsilon$ where $\epsilon$ is proportional to
$\sqrt{\kappa}$. (Note though that this solution
is only valid when $\tau^\epsilon$ is small. It would be interesting to find
an exact solution; we leave this for future work.)

However, we have established our main result: 
de Sitter space, corresponding to $\epsilon=0$ is {\it not} a solution to the 
quantum corrected equations!~ 
Spontaneous symmetry breaking of the de Sitter symmetry
has occurred.

\section{Discussion and Conclusion}

We have argued that de Sitter space is {\it not} a solution to gravity coupled to a cosmological constant when quantum effects 
are taken into account. The classical equations of motion receive quantum corrections
which are singular if the solution is taken to be de Sitter. 
We have argued that the
deviations
from de Sitter are calculable, and 
 that the true solution is deformed away from de Sitter by a 
parameter proportional to $\sqrt{\kappa}$. 

The argument for this was straightforward. In the exact metric of de Sitter space, there are gravitational perturbations
whose propagator is ill defined, and which caused infrared divergences. A small deviation from de Sitter 
parametrized by a small parameter $\epsilon$ allows these
modes to have a well defined propagator. However, the quantum effective action computed around this new metric
now generically has terms which go as $1\over \epsilon$. The quantum equations of motion are singular as we take $\epsilon$ to zero,
and cause de Sitter to not be a solution when quantum corrections are included. 
(Another way to say this is that the de Sitter metric has an infinite action when quantum effects are included.)
The calculation in the previous section argues that these qualitative arguments can be made quantitative, and
that the perturbation away from de Sitter can be computed in perturbation theory.

These arguments are related to previous arguments in the literature, for instance by Polyakov. While Polyakov has argued that
scalar field theory in de Sitter (using the in-out formalism) already leads to an instability, we have shown that gravitons 
produce an instability in the more controlled in-in formalism. The in-in formalism is expected to 
asymptote to the in-out result when the intermediate time is taken to infinity; it would be interesting
to see if this is the case.

The Schwinger-Keldysh propagator for the gravitational perturbations is enhanced by a factor proportional to
${1\over\sqrt{\kappa}}$. This indicates that the gravitational perturbation series, which is normally
in powers of $\kappa$, is modified. A 1-loop diagram with a Schwinger-Keldysh propagator 
now scales as $\sqrt{\kappa}$ and in general, the perturbation series becomes an expansion in
$\sqrt{\kappa}$.

We should also discuss the occasionally thorny issue of gauge invariance. 
It is well known that tadpoles of gravitons are 
not gauge invariant, and so one might wonder about the stautus of the tadpoles we have calculated. 
The resolution is that our intermediate steps have been performed in a fixed gauge, but our final result 
(that de Sitter is unstable) is a gauge invariant statement.
It  is therefore valid in any gauge. Similarly, the deformed metric is not a gauge invariant quantity, but it
has been presented in a particular gauge, and can be transformed to any gauge of choice.

A more subtle issue is the question of observables in quantum gravity. It is often argued that
correlation functions, even for a fixed geodesic distance,
are not well defined; this is roughly because 
any pointlike sources are smeared into black holes.
However, our corrections are of order $\sqrt{\kappa}$ and therefore scale faster than any
perturbative effect in quantum gravity, including the size of black holes. They are hence dominant at weak 
coupling and will not be washed out by
quantum gravity effects.

Finally we note that this solution to the issue of the IR divergences may potentially lead to observable effects, at
least if the Hubble scale is large enough. We leave this issue for future work.

\section{Acknowledgments}

This work was supported in part by  NSF grant PHY-1316792.

\end{document}